\documentclass[conference]{IEEEtran}

\ifCLASSINFOpdf
\else
\fi
\usepackage{graphicx}
\usepackage{bbding}
\usepackage{pifont,geometry,hyperref}
\usepackage{mathrsfs}
\usepackage{multirow}
\usepackage{amssymb,color,subfig}
\usepackage{amsmath,amsfonts,algorithm,algorithmic,epsfig}
\usepackage{cite}
\usepackage{multirow}
\begin{document}
%
\title{Inferring Gene Regulatory Network Using An Evolutionary Multi-Objective Method}

\author{\IEEEauthorblockN{Yu Chen}
\IEEEauthorblockA{School of Science, \\
Wuhan University of Technology,\\
Wuhan, 430070, China\\
Email: ychen@whut.edu.cn}
\and
\IEEEauthorblockN{Xiufen Zou}
\IEEEauthorblockA{School of Mathematics \\ and Statistics, Wuhan University,\\
Wuhan, 430072, China\\
Email: xfzou@whu.edu.cn}
}
\maketitle

\begin{abstract}
Inference of gene regulatory networks (GRNs) based on experimental data is a challenging task in bioinformatics. In this paper, we present a bi-objective minimization model (BoMM) for inference of GRNs, where one objective is the fitting error of derivatives, and the other is the number of connections in the network. To solve the BoMM efficiently, we propose a multi-objective evolutionary algorithm (MOEA), and utilize the separable parameter estimation method (SPEM) decoupling the ordinary differential equation (ODE) system. Then, the Akaike Information Criterion (AIC) is employed to select one inference result from the obtained Pareto set. Taking the S-system as the investigated GRN model, our method can properly identify the topologies and parameter values of benchmark systems. There is no need to preset problem-dependent parameter values to obtain appropriate results, and thus, our method could be applicable to inference of various GRNs models.
\end{abstract}


%
\IEEEpeerreviewmaketitle

\section{Introduction}
Inference of gene regulatory networks (GRNs) is important to understand every detail and principle of biological system, and one of the most popular methods is to reconstruct GRNs by time-course data. In the past decades, the Boolean networks \cite{Davidich2008}, the Bayesian networks \cite{Friedman2000,Werhli2006}, and the ordinary differential equation (ODE)  networks \cite{Tsai2005,Lee2012,Li2015,Xiao2015}, etc., have been proposed to reconstruct GRNs, and meanwhile, the corresponding algorithms were also proposed to infer the network topologies and parameter values.

In this paper, we are devoted to infer the S-system model of GRN is, a popular nonlinear ODE model that represents the dynamic process of biochemical system. It is a nonlinear ODE system
\begin{equation}\label{S-System}
\dfrac{dX_i}{dt}=\alpha_i\prod_{j=1}^NX_{j}^{g_{ij}}-\beta_i\prod_{j=1}^NX_{j}^{h_{ij}},\quad i=1,\dots,N,
\end{equation}
where $X_i$ represents the expression level of gene $i$,  and $N$ is the total number of genes in the investigated network. There are totally $N*(2N+2)$ parameters to be specified in an S-system, where $\alpha_i,\beta_i\in\mathbb{R^+}$ are positive rate constants, and $g_{i,j}$ and $h_{i,j}\in\mathbb{R}$ are kinetic constants.

Generally speaking, it is an inverse problem to infer the S-system using the time course data on expression levels of genes. A general way is to minimize the normalized errors between experimental time course data and numerical results \cite{Tominaga2000} , that is,
\begin{equation}
\min\quad err(\Theta)=\sum_{t=0}^T\sum_{i=1}^N\left(\frac{X_{i,cal}(t)-X_{i,exp}(t)}{X_{i,exp}(t)}\right)^2,
\end{equation}
where $X_{i,cal}(t)$ is the numerical results of $X_i$ at time $t$, and $X_{i,exp}(t)$ refers to the experimental value of gene expression level. However, the inverse problem is usually ill-posed,  and an $L_1$ regularizer could be introduced to penalize the data error  \cite{Palafox2013}. In this way, the minimization model is improved as
\begin{equation}
\min\quad err(\Theta)+\lambda L(\Theta).
\end{equation}
Here,
\begin{equation}
L(\Theta)=\sum_{i=1}^N\sum_{j=1}^N|g_{i,j}+h_{i,j}|
\end{equation}
is the $L_1$ norm of parameter vector $\Theta$, and $\lambda$ is the penalization parameter. However, it is a challenging task to preset an appropriate value of $\lambda$ for GRNs with unknown properties because it is often problem-dependent.

One of the ways to eliminate the difficulty of set appropriate values of $\lambda$ is to convert the single-objective optimization models to bi-objective models. Liu and Wang \cite{Liu2008} proposed a three-objective optimization model simultaneously minimizing the concentration error, slope error and interaction error to find a suitable S-system model structure and its corresponding parameter values, however, they transformed it to a single-objective optimization problem solved by a hybrid differential evolution algorithm, that is, a prior preference was introduced to the inference process. Spieth {\emph et al.} \cite{Spieth2005} took the data error and connectivity number as two minimization objective, but the S-system was integrally inferred because no decoupling method was employed. As a consequence, the obtained Pareto front contained numerous Pareto vectors, and it was challenging to choose the appropriate Pareto solution(s) as the potential S-system setting(s). Koduru \emph{et al.} \cite{Koduru2008} and Cai \emph{et al.} \cite{Cai2012} simultaneously minimized data error for several different data sets, but, did not optimize a metric to make the obtained network sparse.

When  GRNs are reconstructed via evolutionary computation methods, another problem arises with the heavy evaluation process of candidate parameter settings.  To evaluate a set of parameters generated by recombination strategies, the ODE system must be solved via some numerical method, for example, the Runge-Kutta method. However, all equations of a coupled  ODE system must be simultaneously solved, and thus, the evaluation process is computationally heavy. To reduce the complexity of individual evaluation, Tsai and Wang\cite{Tsai2005} used a linear Lagrange polynomial to decouple the ODE system, however, another parameter that has to be regulated was introduced. Liu \emph{et al.}\cite{Liu2012}  developed a separable parameter estimation method (SPEM) to decouple the S-System. By these means, each ODE can be respectively identified by optimization methods.

In order to eliminate the pruning process for setting appropriate parameter values, in this paper we present a multi-objective optimization model for inference of GRNs, and decoupling the S-system  by the SPEM. Then, a multi-objective evolutionary algorithm is proposed to obtain the network topology and parameter values. Although SPEM includes computation process of inverse matrices, our algorithms can perform well for benchmark problems, because only small population is needed to obtain satisfactory results.

The rest of this paper is organized as follows. Section \ref{Sec2} introduces SPEM and bi-objective optimization model for GRN inference, and a proposed multi-objective evolutionary algorithm is introduced in Section \ref{Sec3}. Then, Section \ref{Sec4} performs a preliminary comparison between the obtained Pareto sets with results reported in references. Ultimate inference results obtained by Akaike Information Criterion (AIC) are characterized in Section \ref{Sec5}, and finally, Section \ref{Sec6} draws the conclusions and presents the future work.
\section{Method}\label{Sec2}

\subsection{Decoupling S-systems via the Separable Parameter Estimation Method \cite{Liu2012}}
Trying to fit the curves of derivatives instead of the function curves, the SPEM minimize the difference between ${dX_i}/{dt}$ and $$\hat{\alpha_i}\prod_{j=1}^NX_{j}^{\hat{g_{ij}}}-\hat{\beta_i}\prod_{j=1}^NX_{j}^{\hat{h_{ij}}},$$ where $$\hat{\Theta}=(\hat{\alpha_i},\hat{\beta_i},\hat{g_{i,j}},\hat{h_{i,j}})$$ is a candidate parameter vector of the S-system, and ${dX_i}/{dt}$ is approximated by the five-point numerical derivative method. Then, SPEM constructs the following minimization problem
\begin{eqnarray}\label{SPEM1}
&\min &J_i(\alpha_i ,\beta_i,g_i,h_i)=\nonumber\\
&\sum \limits_{k=1}^n &\left[S_i(t_k)-\left(\alpha_i\prod_{j=1}^NX_{j}^{g_{ij}}-\beta_i\prod_{j=1}^NX_{j}^{h_{ij}} \right)\right],
\end{eqnarray}
where $g_i=(g_{i,1},\dots,g_{i,N})$, $h_i=(h_{i,1},\dots,h_{i,N})$, $S_i(t_k)$ is the approximate value of ${d(X_i)}/{dt}$ at time $t_k$. Denoting $\Gamma_i=(\alpha_i,\beta_i)$, $S_i=[S_i(t_1),\dots,S_i(t_n)]$ and
$$\tilde{X_i}=\left(
                \begin{array}{cc}
                  \prod_{j=1}^NX_{j,t_1}^{g_{ij}}   & -\prod_{j=1}^NX_{j,t_1}^{h_{ij}} \\
                  \vdots & \vdots \\
                  \prod_{j=1}^NX_{j,t_n}^{g_{ij}} & -\prod_{j=1}^NX_{j,t_n}^{h_{ij}} \\
                \end{array}
              \right),
$$
we can rewrite (\ref{SPEM1}) as
\begin{equation}\label{SPEM2}
\min J_i(\Gamma_i,g_i,h_i)=\|S_i-\tilde{X_i}\Gamma_i\|_2^2,
\end{equation}
where $\|\cdot\|$ is the $l_2$-norm. Note that $\tilde{X_i}$ does not depend on $\Gamma_i$, $\Gamma_i$ can be estimated as
\begin{equation}\label{SPEM3}
\hat{\Gamma}_i=\left(\tilde{X}_i^T\tilde{X}_i\right)^{-1}\tilde{X}_i^TS_i.
\end{equation}
Substitute (\ref{SPEM3}) into (\ref{SPEM2}), we can infer each equation of S-system by
\begin{eqnarray}\label{SPEM4}
\min J_i(g_i,h_i)=S_i^T[I-\tilde{X_i}(\tilde{X_i}^T\tilde{X_i})^{-1}\tilde{X_i}^T]S_i. \nonumber\\
i=1,\dots,N
\end{eqnarray}
After the kinetic constants are obtained, corresponding rate constants can be figure out by (\ref{SPEM3}). Then, the coupled S-system is decoupled, and parameters can be estimated equation by equation.
\subsection{The Multi-objection Optimization Model}
To infer parameters values equation by equation, we propose a multi-objective optimization (MO) model
\begin{equation}\label{MO}
               \min \quad \left\{\begin{aligned} & J_i(g_i,h_i) \\
               & \|g_i\|+\|h_i\|.\end{aligned}\right.
             \end{equation}
The first objective minimizes the error for estimation of derivatives, and the second objective is employed to obtain a sparse network, where $\|\cdot\|$ refers to the $l_0$-norm of parameter vectors, i.e., number of connections in the network. Then, we propose a multi-objective evolutionary algorithm to obtain the parameter values.

\section{Multi-objective Evolutionary Algorithm}\label{Sec3}
Because we employ the SPEM decoupling the S-system, we can infer ODE equations one by one. In the proposed  multi-objective evolutionary algorithm (MOEA), each equation is represented by a combination of bit-string and real variables
$$\mathbf x=(\mathbf{bx},\mathbf{rx} ),$$
where  $\mathbf{bx}=(bg_1,\dots,bg_N,bh_1,\dots,bg_{N})$ and $\mathbf{rx}=(g_1,\dots,g_N,h_1,\dots,h_N)$ are respectively bit-strings and real vector of length $2N$. If $bg_i=1$, it means that in the model the parameter $g_i$ is not zero;  If $bh_i=1$, it means that in the model the parameter $h_i$ is not zero. Representing a model via $\mathbf x$ can definitely express the model topology by bit-strings, and consequently, no threshold is needed to round a small parameter value to zero.  The MOEA to solve the multi-objective optimization model (\ref{MO}) is described as follows.
\begin{description}
                 \item[Step1: ]  Randomly generate two populations $[Pop,RPop]$ and $[OldPop,OldRPop]$ of  $PopSize$ individuals. Here $Pop$ and $OldPop$ are the binary populations, and $RPop$ and $OldRPop$ are the real populations. Evaluate $[Pop,RPop]$ and $[OldPop,OldRPop]$ via  (\ref{MO}). Set $[ArchivePop,ArchiveRPop]=[OldPop,OldRPop]$ and $\mathbf{POOL}=RPop$; initialize $\mathbf{gbest}=(\mathbf{bgbest},\mathbf{rgbest})$ to be a randomly selected individual in $[Pop,RPop]$;
                 \item[Step2: ] Generate $Popsize$ offsprings to construct the intermediate population $[IPop,IRPop]$ and evaluate it.
                 \item[Step3: ] Set $[OldPop,OldRPop]=[Pop,RPop]$. Perform non-dominated sorting on the combination of  $[Pop,RPop]$ and $[IPop,IRPop]$. Then, sort it via the dominance rank and values of the second objective in ascending order;
                 \item[Step4: ] Select $PopSize$ best individuals to update the population  $[Pop,RPop]$; update $[ArchivePop,ArchiveRPop]$ and $\mathbf{POOL}$ via $[Pop,RPop]$; randomly select a non-dominated individual as $\mathbf{gbest}$;
                 \item[Step 5: ] If the stopping criterion is not satisfied, goto Step2; otherwise, output the non-dominated solutions and the iteration process ceases.
 \end{description}
 Because the SPEM is employed decoupling the S-system, there are only $2N$ parameters to identify each ODE equation. Note that the second minimization objective as the total number of connections. There are at most  $2N+1$ non-domimnated solutions in the Pareto set. Then, we set the population size greater than $2N+1$, and no diversity mechanisms is needed to obtain a diverse Pareto front. However, if we take the norm of parameter vectors as the second objective that can be any non-negative real value, to obtain a diverse Pareto front a diverse-keeping strategy is necessary. As a consequence, the time complexity increases. Meanwhile, the obtained Pareto set could contain several Pareto solutions with same network topology and similar parameter values, which does not deserve the increased time complexity to some extent.

 Due to the several discrete values of the second objective, the population could be absorbed by some network topology (confirmed by the bit-string $\mathbf{bx}$) that is easy to locate. Thus, we employ a pool of real vectors that keep the diversity of obtained solutions at the early stage of evolution and focus on local exploitation at the late stage.
 \subsection{Generation of Offsprings}
 To generate candidate solutions, three parents are randomly selected. According to the method proposed in \cite{Chen2015}, three parents are selected as follows.
 \begin{itemize}
   \item With a probability $p_1$, two different parents $\mathbf {x_1}=(\mathbf {bx_1},\mathbf {rx_1})$ and $\mathbf{x_2}=(\mathbf {bx_2},\mathbf {rx_2})$ are selected from the present population $[Pop, RPop]$, and $\mathbf {x_3}=(\mathbf {bx_3},\mathbf {rx_3})$ is selected from the last population $[OldPop, OldRPop]$;
   \item otherwise, three different parents $\mathbf {x_1}=(\mathbf {bx_1},\mathbf {rx_1})$, $\mathbf{x_2}=(\mathbf {bx_2},\mathbf {rx_2})$ and $\mathbf{x_3}=(\mathbf {bx_3},\mathbf {rx_3})$ are randomly selected from the Archive $[ArchivePop,ArchiveRPop]$;
 \end{itemize}
Then, compare $\mathbf{x_1}$ with $\mathbf{x_2}$, and initialize the offspring $\mathbf{off}=(\mathbf{boff},\mathbf{roff})$ as the winner.
\subsubsection{Generation of Binary Offsprings}
 To generate the binary offspring $\mathbf{boff}=(boff(1),\dots, boff(N))$, we employ $\mathbf{gbest}=(\mathbf{bgbest},\mathbf{rgbest})$ guiding the search. $\forall i\in\{1,2,\dots,2N\}$,
\begin{itemize}
  \item If $bx_2(i)=bx_3(i)$ and $bx_1(i)\neq gbest(i)$ hold, $boff(i)$ is set to be $gbest(i)$ with probability $1-p_2$;
  \item If $bx_2(i)=bx_3(i)$ and $bx_1(i)= gbest(i)$ hold, $boff(i)$ is randomly generated with probability $p_2$;
  \item otherwise, $boff(i)$ is randomly generated with probability $p_2$.
\end{itemize}
\subsubsection{Generation of Real Offsprings}
The real offspring $\mathbf{roff}=(roff(1),\dots, roff(N))$ is generated as follows.
\begin{itemize}
  \item With probability $p_3$,
  $$roff(i)=rx_1(i)+F\cdot(rx_2(i)-rx_3(i));$$
  \item otherwise,
  $$roff(i)=roff(i)+F\cdot(pool_2(i)-pool_3(i));$$
\end{itemize}
Here $\mathbf{pool_1}=(pool_1(1),\dots,pool_1(N))$ and $\mathbf{pool_2}=(pool_2(1),\dots,pool_2(N))$ are two real vector randomly selected from the real vector set $\mathbf{POOL}$;
\subsection{Update of the Archive}
At the beginning, the archive is randomly initialized. During the evolving process,  $\mathbf{arc_{w}}$, the worst archive member  whose second objective value is greatest  will be replaced once an offspring $\mathbf{off}$ is generated dominating $\mathbf{arc_{w}}$.
\subsection{Update of the Real Pool}
To enhance the global exploration and local exploitation abilities, we employ a real pool $\mathbf{POOL}$ with its corresponding objective vector set $\mathbf{F_{pool}}$ in the proposed MOEA.  $\mathbf{POOL}$ is initialized as the real population $RPop$ at the beginning, and during the evolving process, a randomly selected real vector $\mathbf{pool}\in\mathbf{POOL}$ is updated by $\mathbf{roff}$ if it dominates $\mathbf{pool}$ via the corresponding objective vectors.

\section{Preliminary Results for Inference of Benchmark S-Systems}\label{Sec4}
\subsection{Inference Results of Benchmark S-Systems Obtained via Sole Data Set}
To evaluate the performance of our proposed method, we utilize it inferring four benchmark S-systems. For each benchmark S-system, we employ a population of size 20 to solve the proposed multi-objective optimization model (\ref{MO}). After 4000 iterations, the obtained non-dominated solutions and their objective values are saved. The parameter settings listed in Tab. \ref{Tab5} are controlled via a process indicator $$I=0.9-\left\lfloor \frac{Gen}{MaxGen/10}\right\rfloor, $$ where $Gen$ is the generation counter, $MaxGen$ is the maximum generation  for the stopping criterion.   We perform 20 independent runs for each ODE equation of four benchmark systems, and the best obtained results with smallest data errors are respectively included in Tabs. \ref{Tab1}, \ref{Tab2}, \ref{Tab3} and \ref{Tab4}. Both precise network parameters and obtained results are listed in the tables, where the inference results are enclosed by parentheses. For convenience of comparison, the first number in the parentheses is results obtained by our method, and the second is what was reported in references. If the inference result for a run cannot correctly identify all connections, we take it as a failure run of the proposed MOEA. Then, the success rates are also included in the last columns of the tables.
\begin{table}
\caption{Parameter Settings in the Proposed MOEA.}\label{Tab5}
\centering
\begin{tabular}{|c|l|}
  \hline
  $Parameter$ & Parameter setting\\
  \hline
  $p_1$ & $p_1=4\times(I-0.5)^2$ \\
  \hline
  $p_2$ & $p_2=0.2$ \\
  \hline
  $p_3$& $p_3=\exp{(-I)}$ \\
  \hline
  $F$ & $F=(I+0.1)\times rand(-1,1)$\\
  \hline
\end{tabular}
\end{table}
\subsubsection{Inferring Results of {\bf S1}}
\begin{figure}[!t]
     \centering
     \includegraphics[width=2in]{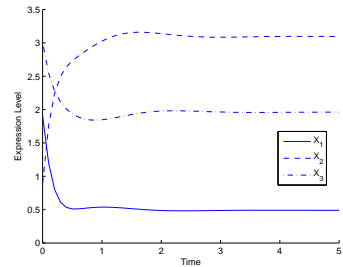}%
     \caption{Trajectories of variables in {\bf S1}.  }
     \label{Fig1}
     \end{figure}

The first benchmark system {\bf S1} is a 3-D system where the parameters are listed in Tab.\ref{Tab1}. For initial conditions $X_1(0)=1.9$, $X_2(0)=0.9$ and $X_3(0)=3.0$, data are sampled from time 0 to 5 with stepsize 0.1, and  the time-course curves are illustrated in Fig.\ref{Fig1} \cite{Liu2012}.  Correct network connections for $X_2$ and $X_3$ are always included in the obtained non-dominated sets, however, for the first equation, 3 of 20 runs cannot correctly identify the network topology. Curve of $X_1$ in Fig. \ref{Fig1} demonstrates that derivative of $X_1$ quickly decrease to zero after around 0.5 time unit, which implies that only five sample point of derivative are not zero. Because our method is based on fitting of derivatives, it cannot perform well for this case.

 \begin{table*} \tiny
    \centering
    \caption{Precise and Obtained Values for Parameters of Benchmark System {\bf S1}.}
    \label{Tab1}
    \begin{tabular}{|c|c|c|c|c|c|c|c|c|c|}
      \hline
      $i$ & $\alpha_{i}$ & $g_{i1}$ & $g_{i2}$ & $g_{i3}$ & $\beta_{i}$ & $h_{i1}$ & $h_{i2}$ & $h_{i3}$ & success rate\\
      \hline

      \multirow{2}{*}{1}& 12 & 0 & 0 & -0.8 & 10 & 0.5 & 0 & 0 & \multirow{2}{*}{85\%} \\
      &(12.00,10.97) & (0.00,0.00) & (0.00,0.00) & (-0.80,-0.97) & (9.98,8.81) & (0.50,0.60) & (0.00,0.00) & (0.00,0.00) &  \\
      \hline
      \multirow{2}{*}{2}& 10 & 0.5 & 0 & 0& 3& 0(0) & 0.75 & 0 & \multirow{2}{*}{100\%}\\
      &(9.80,9.01) & (0.51,0.59)& (0.00,0.00) & (0.00,0.00)  & (2.83,2.17) & (0.00,0.00)  & (0.78,0.89)& (0.00,0.00) &  \\
      \hline
      \multirow{2}{*}{3} & 3 & 0& 0.75 & 0& 5 & 0& 0 & 0.5  &\multirow{2}{*}{100\%}\\
      &(2.56,2.94) & (0.00,0.00) & (0.82,0.76)& (0.00,0.00)  & (4.5,4.92)& (0.00,0.00)  & (0.00,0.00)& (0.55,0.50)&  \\
      \hline
    \end{tabular}
    \end{table*}
\subsubsection{Inferring Results of {\bf S2}}
\begin{figure}[!t]
     \centering
     \includegraphics[width=2in]{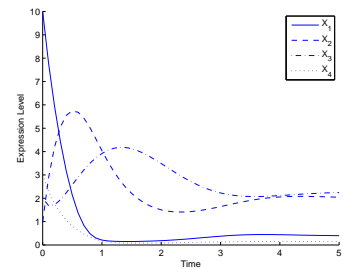}%
     \caption{Trajectories of variables in \bf{S2}. }
     \label{Fig2}
     \end{figure}

The second benchmark system is a 4-D system where the parameters are listed in Tab.\ref{Tab2}. For initial conditions $X_1(0)=10$, $X_2(0)=1$,$X_3(0)=2$ and $X_4(0)=3$, data are sampled from time 0 to 5 with stepsize 0.1, and the time-course curves are illustrated in Fig.\ref{Fig2} \cite{Liu2012}. Compared with results presented in \cite{Liu2012}, the precisions of parameter values are a bit lower. However, we can obtained the correct  network topology with 100\% success rate except for the third equation. Considering that in \cite{Liu2012} the initial parameter setting is generated by normal mutation performed on the known parameter values,  we can conclude that our method is competitive to the method presented in \cite{Liu2012} because we randomly generate initial parameters at the beginning.

 \begin{table*} \tiny
    \centering
    \caption{Precise and Obtained Values for Parameters of Benchmark System {\bf S2}. }
    \label{Tab2}
    \begin{tabular}{|c|c|c|c|c|c|c|c|c|c|c|c|}
      \hline
      $i$ & $\alpha_{i}$ & $g_{i1}$ & $g_{i2}$ & $g_{i3}$ & $g_{i4}$ & $\beta_{i}$ & $h_{i1}$ & $h_{i2}$ & $h_{i3}$& $h_{i4}$ & success rate\\
      \hline

      \multirow{2}{*}{1} & 12 & 0 & 0 & -0.8 & 0 & 10 & 0.5 & 0 & 0 & 0 &\multirow{2}{*}{100\%} \\
      & (11.80,12.00) & (0.00,0.00) & (0.00,0.00) & (-0.79,-0.80) & (0.00,0.00) & (10.00,9.97) & (0.50,0.50) & (0.00,0.00) &  (0.00,0.00) & (0.00,0.00) & -\\
      \hline
      \multirow{2}{*}{1} & 8 & 0.5& 0& 0 & 0& 3 & 0& 0.75 & 0 & 0 & \multirow{2}{*}{100\%}\\
      & (8.00,8.00) & (0.50,0.50) & (0.00,0.00) & (0.00,0.00) & (0.00,0.00) & (3.00,3.01)  & (0.00,0.00)  & (0.75,0.75) & (0.00,0.00) & (0.00,0.00) &  \\
      \hline
      \multirow{2}{*}{3}& 3 & 0& 0.75 & 0& 0& 5 & 0& 0 & 0.5 & 0.2 &\multirow{2}{*}{90\%}\\
      & (3.00,3.00) & (0.00,0.00) & (0.76,0.75) & (0.00,0.00) & (0.00,0.00) & (5.10,5.01)  & (0.00,0.00)  & (0.00,0.00) & (0.51,0.50)  & (0.20,0.20) &  \\
      \hline
      \multirow{2}{*}{4} & 2 & 0.5 & 0& 0& 0& 6& 0& 0& 0 & 0.8 &\multirow{2}{*}{100\%}\\
      & (1.90,2.00) & (0.49,0.50) & (0.00,0.00) & (0.00,0.00) & (0.00,0.00) & (5.80,6.00)  & (0.00,0.00)  & (0.00,0.00) & (0.00,0.00) & (0.80,0.80) & \\
      \hline
    \end{tabular}
    \end{table*}

\subsubsection{Inferring Results of {\bf S3}}
\begin{figure}[!t]
     \centering
     \includegraphics[width=2in]{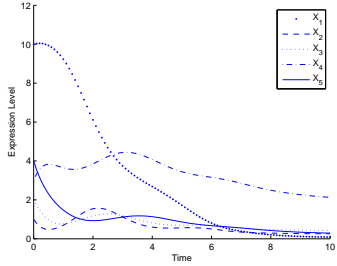}%
     \caption{Trajectories of variables in \bf{S3}.}
     \label{Fig3}
     \end{figure}

The third benchmark system is a 5-D system where the parameters are listed in Tab.\ref{Tab3}. For initial conditions $X_1(0)=10$, $X_2(0)=1$,$X_3(0)=2$, $X_4(0)=3$ and $X_4(0)=4$, data are sampled from time 0 to 10 with stepsize 0.1, and the time-course curves are illustrated in Fig.\ref{Fig3} \cite{Liu2012}. For this system, precisions of our obtained results are also lower than those reported in \cite{Liu2012}, especially for the second equation. This is because Liu \emph{et al.} employed a local search procedure to refine the parameter values, while in our method no local searching process is implemented. Meanwhile, because derivatives of $X_2$ fluctuate in a small interval, our method only successfully obtain the correct connection of the second equation with success rate 85\%.

\begin{table*}\tiny
    \centering
    \caption{Precise and Obtained Values for Parameters of Benchmark System {\bf S3}.}
    \label{Tab3}
    \begin{tabular}{|c|c|c|c|c|c|c|c|c|c|c|c|c|c|}
      \hline
      $i$ & $\alpha_{i}$ & $g_{i1}$ & $g_{i2}$ & $g_{i3}$ & $g_{i4}$ & $g_{i5}$ & $\beta_{i}$ & $h_{i1}$ & $h_{i2}$ & $h_{i3}$ & $h_{i4}$ & $h_{i5}$ & success rate \\
      \hline
      \multirow{2}{*}{1} & 2 & 0 & 0& 0 & 0 & 0 & 2 & 0.5 & 0 & 0 & 0 & -1 & \multirow{2}{*}{100\%}\\
      & (1.99,2.00) & (0.00,0.00) & (0.00,0.00)  &  (0.00,0.00) & (0.00,0.00) & (0.00,0.00) & (1.99,2.00)  &  (0.50,0.50) & (0.00,0.00) & (0.00,0.00) & (0.00,0.00) &  (-1.00,-1.00) & \\
      \hline
      \multirow{2}{*}{2} & 2 & 0.5 & 0& 0& 0 & -1& 4 & 0& 0.5 & 0 & 0 & 0&\multirow{2}{*}{10\%}\\
      & (1.86,1.01) & (0.52,0.99) & (0.00,0.00)  &  (0.00,0.00) & (0.00,0.00) & (-1.03,-1.00) & (3.8351,4.03)  &  (0.00,0.00) & (0.52,0.50) & (0.00,0.00) & (0.00,0.00) &  (0.00,0.00) & \\
      \hline
      \multirow{2}{*}{3} & 4(3.9) & 0(0) & 0.5(0.51) & 0(0) & 0(0) & 0(0) & 4(3.9) & 0(0) & 0(0) & 0.8(0.81) & 0(0) & 0(0) & \multirow{2}{*}{85\%}\\
       & (3.93,4.00) & (0.00,0.00) & (0.51,0.50)  &  (0.00,0.00) & (0.00,0.00) & (0.00,0.00) & (3.93,4.00)  &  (0.00,0.00) & (0.00,0.00) & (0.81,0.80) & (0.00,0.00) &  (0.00,0.00) & \\
       \hline
      \multirow{2}{*}{4} & 4 & 0 & 0 & 0.8 & 0 & 0 & 1 & 0 & 0 & 0 & 1 & 0 & \multirow{2}{*}{100\%}\\
      & (3.97,4.00) & (0.00,0.00) & (0.00,0.00)  &  (0.81,0.80) & (0.00,0.00) & (0.00,0.00) & (0.98,1.00)  &  (0.00,0.00) & (0.00,0.00) & (0.00,0.00) & (1.01,1.00) &  (0.00,0.00) &  \\
      \hline
      \multirow{2}{*}{5} & 1& 0& 0& 0 & 1 & 0& 4 & 0& 0& 0 & 0 & 0.5 & \multirow{2}{*}{100\%}\\
      & (0.98,1.00) & (0.00,0.00) & (0.00,0.00)  &  (0.00,0.00) & (1.00,1.00) & (0.00,0.00) & (3.97,4.00)  &  (0.00,0.00) & (0.00,0.00) & (0.00,0.00) & (0.00,0.00) &  (0.50,0.50) &  \\
      \hline
    \end{tabular}
    \end{table*}

\subsubsection{Inferring Results of {\bf S4}}
\begin{figure}[!t]
     \centering
     \includegraphics[width=2in]{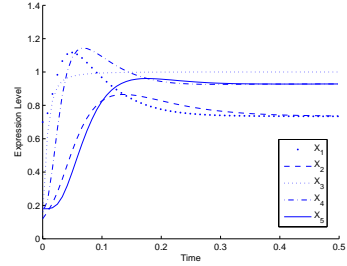}%
     \caption{Trajectories of variables in \bf{S4}.}
     \label{Fig4}
     \end{figure}

The fourth benchmark system is a 5-D system where the parameters are listed in Tab.\ref{Tab4}\cite{Palafox2013}. To compare with the results reported in \cite{Palafox2013}, we employ a data set of 60 samples generated from time 0 to 0.5 with stepsize 0.0083 with initial conditions $X_1(0)=0.7$, $X_2(0)=0.12$,$X_3(0)=0.14$, $X_4(0)=0.16$ and $X_4(0)=0.18$, and the time-course curves are illustrated in Fig.\ref{Fig4}. Results listed in Tab. \ref{Tab4} demonstrate that for equations 1, 2,and 5, the best obtained parameter values are more precise than what were reported in \cite{Palafox2013}, however, inference results for equations 3 and 4 are not satisfactory.  Fig. \ref{Fig4} illustrates that the curves of $X_3$ and  $X_4$ rapidly reach equilibrium states after short increasing procedure, and consequently, our method, based on fitting of derivatives (slopes), could not identify the network connections well.

    \begin{table*}\tiny
    \centering
    \caption{Precise and Obtained Values for Parameters of Benchmark System {\bf S4}. }
    \label{Tab4}
    \begin{tabular}{|c|c|c|c|c|c|c|c|c|c|c|c|c|c|}
      \hline
      $i$ & $\alpha_{i}$ & $g_{i1}$ & $g_{i2}$ & $g_{i3}$ & $g_{i4}$ & $g_{i5}$ & $\beta_{i}$ & $h_{i1}$ & $h_{i2}$ & $h_{i3}$ & $h_{i4}$ & $h_{i5}$ & success rate\\
      \hline
      \multirow{2}{*}{1} & 5 & 0 & 0 & 1 & 0 & -1 & 10 & 2 & 0 & 0 & 0 & 0 & \multirow{2}{*}{85\%} \\
      & (4.91, 4.38) & (0.00, 0.00) & (0.00, 0.00) & (1.00, 1.43) & (0.00, 0.00)  & (-1.01, -0.90) & (9.89, 9.57) & (2.02, 1.47) & (0.00, 0.00)  &(0.00, 0.00)  &(0.00, 0.00)  &(0.00, 0.00) & \\
      \hline
      \multirow{2}{*}{2} & 10 & 2 & 0 & 0& 0& 0& 10 & 0 & 2 & 0 & 0 & 0  & \multirow{2}{*}{100\%}\\
      & (9.99, 9.32) & (2.00, 1.79) & (0.00, 0.00) & (0.00, 0.00) & (0.00, 0.00)  & (0.00, 0.00) & (9.99, 10.56) & (0.00, 0.00) & (2.00, 2.06)  &(0.00, 0.00)  &(0.00, 0.00)  &(0.00, 0.00) & \\
      \hline
      \multirow{2}{*}{3} & 10& 0 & -1& 0 & 0& 0& 10 & 0 & -1& 2& 0 & 0& \multirow{2}{*}{10\%}\\
      & (9.47, 10.88) & (0.00, 0.00) & (-1.05, -1.66) & (0.00, 0.00) & (0.00, 0.00)  & (0.00, 0.00) & (9.49, 9.85) & (0.00, 0.00) & (-1.04, -1.25)  &(1.84, 1.88)  &(0.00, 0.00)  &(0.00, 0.00) & \\
      \hline
      \multirow{2}{*}{4} & 8 & 0 & 0 & 2 & 0& -1 & 10 & 0& 0& 0 & 2 & 0 & \multirow{2}{*}{90\%}\\
      & (22.37, 8.33) & (0.00, 0.00) & (0.00, 0.00) & (1.89, 1.90) & (0.00, 0.00)  & (-0.54, -0.75) & (24.91, 9.89) & (0.00, 0.00) & (0.00, 0.00)  &(0.00, 0.00)  &(0.93, 2.24)  &(0.00, 0.00) & \\
      \hline
      \multirow{2}{*}{5} & 10 & 0(0) & 0 & 0 & 2 & 0 & 10 & 0 & 0 & 0 & 0 & 2 & \multirow{2}{*}{100\%}\\
      & (9.92, 9.63) & (0.00, 0.00) & (0.00, 0.00) & (0.00, 0.00) & (2.06, 2.06)  & (0.00, 0.00) & (9.89, 10.66) &  (0.00, 0.00) &  (0.00, 0.00)  & (0.00, 0.00)  & (0.00, 0.00)  &(2.00, 2.14) & \\
      \hline
    \end{tabular}
    \end{table*}

\subsection{Improving Inference Results by Multiple Data Sets}
To improve the inference results of our method, we employ multiple data sets inferring the S-system {\bf S4}. For comparison with results reported in \cite{Palafox2013}, we generate four data set with initial conditions
\begin{enumerate}
\item $X_1(0)=0.7$, $X_2(0)=0.12$,$X_3(0)=0.14$, $X_4(0)=0.16$ and $X_4(0)=0.18$;
\item $X_1(0)=0.7$, $X_2(0)=0.12$, $X_3(0)=0.14$, $X_4(0)=0.16$ and $X_4(0)=0.18$;
\item $X_1(0)=0.1$, $X_2(0)=0.12$, $X_3(0)=0.7$, $X_4(0)=0.16$ and $X_4(0)=0.18$;
\item $X_1(0)=0.1$, $X_2(0)=0.12$,$X_3(0)=0.14$, $X_4(0)=0.16$ and $X_4(0)=0.7$,
\end{enumerate}
where each data set contains 15 data samples. The obtained results are reported in Tab. \ref{Tab6}.
\begin{table*}\tiny
    \centering
    \caption{Improved Inference Results of Benchmark System {\bf S4} by Four Data Sets.}
    \label{Tab6}
    \begin{tabular}{|c|c|c|c|c|c|c|c|c|c|c|c|c|c|}
      \hline
      $i$ & $\alpha_{i}$ & $g_{i1}$ & $g_{i2}$ & $g_{i3}$ & $g_{i4}$ & $g_{i5}$ & $\beta_{i}$ & $h_{i1}$ & $h_{i2}$ & $h_{i3}$ & $h_{i4}$ & $h_{i5}$ & success rate\\
      \hline
      \multirow{2}{*}{1} & 5 & 0 & 0 & 1 & 0 & -1 & 10 & 2 & 0 & 0 & 0 & 0 & \multirow{2}{*}{100\%} \\
      & (4.84, 4.38) & (0.00, 0.00) & (0.00, 0.00) & (1.01, 1.43) & (0.00, 0.00)  & (-1.05, -0.90) & (9.88, 9.57) & (2.03, 1.47) & (0.00, 0.00)  &(0.00, 0.00)  &(0.00, 0.00)  &(0.00, 0.00) & \\
      \hline
     \multirow{2}{*}{2} & 10 & 2 & 0 & 0& 0& 0& 10 & 0 & 2 & 0 & 0 & 0  & \multirow{2}{*}{100\%}\\
      & (9.99, 9.32) & (2.00, 1.79) & (0.00, 0.00) & (0.00, 0.00) & (0.00, 0.00)  & (0.00, 0.00) & (9.99, 10.56) & (0.00, 0.00) & (2.00, 2.06)  &(0.00, 0.00)  &(0.00, 0.00)  &(0.00, 0.00) & \\
      \hline
      \multirow{2}{*}{3} & 10 & 0 & -1 & 0 & 0 & 0 & 10 & 0 & -1 & 2 & 0 & 0 & \multirow{2}{*}{100\%}\\
      & (10.92, 10.88) & (0.00, 0.00) & (-1.00, -1.659) & (0.00, 0.00) & (0.00, 0.00)  & (0.00, 0.00) & (10.92, 9.85) & (0.00, 0.00) & (-1.01, -1.25)  &(1.67, 1.88)  &(0.00, 0.00)  &(0.00, 0.00) & \\
      \hline
      \multirow{2}{*}{4} & 8 & 0 & 0 & 2 & 0& -1 & 10 & 0& 0& 0 & 2 & 0 & \multirow{2}{*}{100\%}\\
      & (8.66, 8.33) & (0.00, 0.00) & (0.00, 0.00) & (1.99, 1.90) & (0.00, 0.00)  & (-0.97, -0.75) & (10.84, 9.89) & (0.00, 0.00) & (0.00, 0.00)  &(0.00, 0.00)  &(1.74, 2.24)  &(0.00, 0.00) & \\
      \hline
      \multirow{2}{*}{5} & 10 & 0 & 0 & 0 & 2 & 0 & 10 & 0 & 0 & 0 & 0 & 2 & \multirow{2}{*}{100\%}\\
      & (9.97, 9.63) & (0.00, 0.00) & (0.00, 0.00) & (0.00, 0.00) & (2.00, 2.06)  & (0.00, 0.00) & (9.97, 10.66) &  (0.00, 0.00) &  (0.00, 0.00)  & (0.00, 0.00)  & (0.00, 0.00)  &(2.01, 2.14) & \\
      \hline
    \end{tabular}
    \end{table*}

By employing four data sets generated by different initial conditions, our method can perfectly identify the network topology of {\bf S4} with success rate 100\%, and the obtained kinetic constant values are generally better than the results reported in \cite{Palafox2013}. However, sometimes rate constants obtained by our method are a bit worse than the compared results, because the rate constants are generated via the Least Square Method (LSM) in our method.

\section{Obtaining the GRN Settings from Candidate Non-dominated Solutions}\label{Sec5}
Then, we select the Pareto solutions with minimum Akaike Information Criterion (AIC) value as the ultimate GRN inference results. The AIC for a candidate Pareto solution is computed as
\begin{equation}
AIC(\mathbf{x})=\log\left(\frac{1}{M}J_i(g_i,h_i)\right)+\frac{2(\|g_i\|+\|h_i\|)}{M},
\end{equation}
where $M$ is the number of data samples \cite{Yao2011}. Based on the results selected from obtained Pareto sets of 20 independent runs, we compute the sensitivity $S_n$ and specificity $S_p$ as follows:
\begin{equation}
S_n=\frac{TP}{TP+FN}
\end{equation}
\begin{equation}
S_p=\frac{TN}{TN+FP}
\end{equation}
where TP, FN, TN and FP represent the true positive (TP), false negative (FN), true negative (TN) and false positive (FP) predictions of the parameters. Average values of $S_n$ and $S_p$ for 20 independent runs for $S_4$ are listed in Tab. \ref{Tab7}.
\begin{table}
    \centering
    \caption{Average Index Values of Inference Results of {\bf S4} Obtained by Four Data Sets.}
    \label{Tab7}
    \begin{tabular}{|c|c|c|c|c|c|c|}
      \hline i & TP & FN &TN &FP & $S_n$ & $S_p$\\
      \hline
      1 & 3 & 0 &6.95 & 0.05 & $1$ & 0.99\\
      \hline
      2 & 2 & 0 &8 &0 & $1$ & 1\\
      \hline
      3 & 3 & 0 & 6.6 &0.4 & 1 & 0.94\\
      \hline
      4 & 3 & 0.05 &6.1 &0.85 & 0.98 & 0.88\\
      \hline
      5 & 2 & 0 &8 &0 & 1& 1\\
      \hline
    \end{tabular}
    \end{table}
It is shown that the sensitivity $S_n$ is equal to 1 except for the fourth equation ($S_n=0.98$), which means that for all equations the true network connections can be identified with a probability approximately equal to 1; however, sometimes the specificity $S_n$ is less than 1, which occurs when the true network topologies plus false positive connections can achieve better fitting errors.
\section{Conclusion}\label{Sec6}
In this paper, we presents an evolutionary multi-objective approach to inference network of S-system, using the SPEM decoupling the ODE system. Then, AIC is employed to select one Pareto solution as the final inference result. Due to the proposed multi-objective optimization model, there is no need to preset any parameter value before the inference MOEA is run. So, our method could be generally applicable to various kind of GRN model. However, we also find that this method is sensitive to the curve of expression level, and multiple data sets are needed to improve its performances. Future work will be focused on inference of GRN based on noisy data, and how to generalize the proposed method to infer big-scale GRNs.

\section*{Acknowledgment}
This work is partially supported by ``the  Fundamental Research Funds for the Central Universities (WUT: 2014-Ia-007)'' and the National Science Foundation of China under grants 61303028, 61173060 and 91230118.

\end{document}